\documentstyle[12pt,axodraw]{article}

\advance\textheight by \topskip
\renewcommand{\baselinestretch}{1.2}
\renewcommand{\thefootnote}{\fnsymbol{footnote}}
\newcommand{\beq}{\begin{equation}}
\newcommand{\eeq}{\end{equation}}
\newcommand{\bea}{\begin{eqnarray}}
\newcommand{\eea}{\end{eqnarray}}
\renewcommand{\(}{\left(}
\renewcommand{\)}{\right)}
\renewcommand{\[}{\left[}
\renewcommand{\]}{\right]}
\renewcommand{\bar}[1]{\overline{#1}}
\def\slash#1{#1\!\!\!/\!\,\,}

\renewcommand{\Im}{\mbox{Im}}
\renewcommand{\Re}{\mbox{Re}}
\newcommand{\D}{{\cal D}}

\begin{document}
\newtheorem{fig}[figure]{Figure}


\begin{titlepage}
\renewcommand{\baselinestretch}{1}
\renewcommand{\thefootnote}{\alph{footnote}}

\thispagestyle{empty}

\vspace*{-1.3cm}
{\bf \hfill                                              LMU--05/96
}\vspace*{-0.3cm}

{\bf \hfill\hfill                                          May 1996
}\vspace*{1.5cm}
{\Large\bf
\begin{center}         Family Structure from Periodic Solutions of
\\                          an Improved Gap Equation
\end{center}}
\vspace*{1cm}
{\begin{center} 
{\large\sc
                           Andreas Blumhofer\footnote{
\makebox[1.cm]{Email:} Blumhofer@Photon.HEP.Physik.Uni-Muenchen.DE}
and
                           Marcus Hutter\footnote{
\makebox[1.cm]{Email:} MH@HEP.Physik.Uni-Muenchen.DE}}
\end{center} }
\vspace*{0cm} {\it \begin{center}
Sektion Physik \\ Ludwig--Maximilians--Universit\"at
M\"unchen \\ Theresienstr.37 \\ D--80333 M\"unchen, Germany
\end{center} } 
\vspace*{2.3cm}

\renewcommand{\thefootnote}{}
\footnotesep0.5cm
\footnote{This work is in part supported by the ``German Israeli
Foundation'' No.I 0304-120.07/93.}
\footnotesep0cm

{\Large \bf \begin{center} Abstract \end{center} }
Fermion mass models usually contain a horizontal symmetry and therefore 
fail to predict the exponential mass spectrum of the Standard Model in a 
natural way.
In dynamical symmetry breaking there are different concepts to introduce a
fermion mass spectrum, which automatically has the desired hierarchy.
In constructing a specific model we show that in some modified gap 
equations periodic solutions with several fermion poles appear. 
The stability of these excitations and the application of this toy model 
are discussed.
The mass ratios turn out to be  approximately $e^\pi$ and $e^{2\pi}$.
Thus the model explains the large ratios of fermion masses between 
successive generations in the Standard Model without introducing large
or small numbers by hand.

\renewcommand{\baselinestretch}{1.2}
\end{titlepage}

\newpage
\renewcommand{\thefootnote}{\arabic{footnote}}
\setcounter{footnote}{0}


\section{Introduction}

Although the Standard Model describes the low energy physics with 
relatively few parameters, model building tends to increase the number of 
particles and the degrees of freedom with a low reduction of the number of
parameters. That means either that the low energy physics can only be 
understood at the Planck scale or that a complex dynamical structure is
responsible for the pattern of masses and couplings.
The family duplication is the central problem and one will not be able to
find the origin of mass generation without facing that problem. 
There were a lot of attempts to understand the mass splittings and the
mixing angles in the CKM--matrix \cite{Stech}-\cite{Foot}.
One possibility is to study different textures as it was done in the 
classical papers by B.Stech and H.Fritzsch \cite{Stech}.
They are followed by a lot of papers partly explaining the high top mass
\cite{Fritzsch}. Some authors have searched for infrared fixed points
of the running couplings to determine the masses and mixing angles 
\cite{Pendleton}. In constructing underlying models predicting the
parameters from first principles one has studied composite models of quarks
and leptons \cite{Katsumata} or has established a horizontal symmetry to
explain the different generations \cite{Foot}.

In all these models a mass spectrum
\bea
        m_k\approx m_1\cdot e^{(k-1)\alpha}\; ,\quad
        \mbox{Generation}\;\; k=1,2,3
\eea       
as it is realized in the Standard Model is hard to describe\footnote{
The masses at a fixed scale (e.g. 1GeV) must be compared rather than the 
on--shell masses as long as the running of the quark masses is not 
involved in the model.}
(see fig.\ref{spectrum} \cite{Koide}). 

\begin{figure}[htb]
\begin{center}
\begin{minipage}{7.5cm}
\begin{center}
\begin{picture}(150,200)
\put(0,0){\vector(0,1){200}}
\put(0,0){\vector(1,0){150}}
\multiput(-2,0)(0,25){8}{\line(1,0){4}}
\multiput(20,-2)(50,0){3}{\line(0,1){4}}
\put(-4,0){\makebox(0,0)[r]{$10^{-4}$}}
\put(-4,25){\makebox(0,0)[r]{$10^{-3}$}}
\put(-4,50){\makebox(0,0)[r]{$10^{-2}$}}
\put(-4,75){\makebox(0,0)[r]{$10^{-1}$}}
\put(-4,100){\makebox(0,0)[r]{$1$}}
\put(-4,125){\makebox(0,0)[r]{$10^{1}$}}
\put(-4,150){\makebox(0,0)[r]{$10^{2}$}}
\put(-4,175){\makebox(0,0)[r]{$10^{3}$}}
\put(-4,200){\makebox(0,0)[r]{$m$($\mu=$1GeV)}}
\put(-25,187){\makebox(0,0)[r]{[GeV]}}
\put(20,-4){\makebox(0,0)[t]{$1$}}
\put(70,-4){\makebox(0,0)[t]{$2$}}
\put(120,-4){\makebox(0,0)[t]{$3$}}
\put(150,-4){\makebox(0,0)[tl]{Generation $k$}}
\put(20,18){\circle*{3}}   \put(20,14){\makebox(0,0)[t]{$e$}}
\put(70,76){\circle*{3}}   \put(70,72){\makebox(0,0)[t]{$\mu$}}
\put(120,106){\circle*{3}} \put(120,102){\makebox(0,0)[t]{$\tau$}}
\put(20,50){\circle*{3}}   \put(20,54){\makebox(0,0)[b]{$d$}}
\put(70,82){\circle*{3}}   \put(70,86){\makebox(0,0)[b]{$s$}}
\put(120,121){\circle*{3}} \put(120,125){\makebox(0,0)[b]{$b$}}
\put(20,44){\circle*{3}}   \put(20,40){\makebox(0,0)[t]{$u$}}
\put(70,104){\circle*{3}}  \put(70,110){\makebox(0,0)[b]{$c$}}
\put(120,166){\circle*{3}} \put(120,170){\makebox(0,0)[b]{$t$}}
\Line(20,18)(70,76)    \Line(70,76)(120,106)
\Line(20,50)(70,82)    \Line(70,82)(120,121)
\Line(20,44)(70,104)   \Line(70,104)(120,166)
\end{picture}
\end{center}
\begin{fig}
  Fermion mass spectrum.
  \label{spectrum}
\end{fig}
\end{minipage}
\end{center}
\end{figure}

E.g. a horizontal symmetry must be broken drastically.
Hence it seems more reasonable to start with a theory which predicts an
exponential mass spectrum in a natural way from the very beginning.

Dynamical symmetry breaking models are excellent candidates.
There are two possibilities to generate excitations. One may interpret 
either the different solutions of a gap equation or the different 
propagator poles of one solution as particles of higher generations. In 
the first case \cite{Triantaphyllou} each solution of the gap equation 
represents another minimum of the effective action. The generation number 
of the particle depends
on the vacuum, where it is living. Hence the excited states must move in
vacuum bubbles, which unfortunately are rather unstable \cite{Anderson}. 
Nevertheless the mass matrices and therefore the CKM matrix 
might in principle be determined by calculating the instanton transitions 
between different vacua and one gets the Standard Model as an effective 
description of that scenario. Hence the model might be phenomenological 
acceptable, if the problem of the bubble instability can be solved. 

In this paper we will concentrate on the second case, where we are 
interested in gap equations with periodic solutions. 
Here the fermion propagator $ S(p^2)=\frac{i}{p\!\!/-B(p^2)} $ with
the dynamically generated mass function $B(p^2)$ should have several poles 
$m_k^2$ with $m_k^2-B(m_k^2)^2=0$, where $m_k$ is the corresponding
mass spectrum. 
This is rather non--trivial since a real continuous periodic function 
$p^2-B(p^2)^2$ cuts the $p^2$--axis both from above and below, which 
alternately leads to poles with positive and negative residue corresponding
to particles and ghosts.
A ghost pole in $S(p^2)$ is avoided, if $B(p^2)^2$ admits a pole or an 
imaginary part in the $p^2$--region between two particle states. 
An imaginary part above a fermion pole appears anyway. It describes the 
decay of the off--shell fermion. But it is unusual that this imaginary part
disappears again at the next fermion pole.

As shown in section \ref{gap} a periodic fermion mass function cannot be
found 
in the simple ladder approximation. One has to improve the corresponding 
gap equation by introducing a running coupling to find periodic solutions
(section \ref{per}). The different poles of the propagator yield mass
ratios consistent with the Standard Model mass spectrum.

In section \ref{mul} we show that the different fermion poles really can be
interpreted as different particle states with own propagators. At the tree
level one cannot decide whether there is one propagator with several poles
or several propagators.

In section \ref{eff} an effective description of our model is given, where 
one naively expects flavor changing transitions. But
using only gauge invariance we will show that the particle spectrum is
stable. 

Section \ref{real} generalizes the discussion of the previous sections
to more realistic models containing a fermion isodoublet per generation.

The model therefore naturally contains the essential properties of the 
mass generation and the mass spectrum of the Standard Model.


\section{The gap equation in the ladder approximation}   \label{gap}

We study a toy model of one fermion field coupled minimally to a boson 
field. In section \ref{real} the fermion field will be transfered 
to the Standard Model
fermions and the boson field to Standard Model gauge bosons or to  
a hidden sector of strongly coupled vector fields.
A fundamental Higgs boson is absent. The dynamical mass generation 
should rather be compared to a QCD--like scenario. 
The fermion selfenergy can be determined by the Schwinger--Dyson
selfconsistency equation 

\begin{minipage}{14.9cm}
\begin{center}
\begin{picture}(230,40)(0,0)
\ArrowLine(20,10)(40,10)        \ArrowLine(50,10)(70,10)
\GOval(45,10)(5,8)(0){0.2}
\Text(95,10)[c]{=}
\ArrowLine(120,10)(145,10)      \ArrowLine(145,10)(195,10)
\ArrowLine(195,10)(220,10)      \DashCArc(170,10)(25,0,180) 4 
\GOval(170,10)(5,8)(0){0.7}     \GOval(195,10)(5,5)(0){0.2}
\end{picture}
\end{center}
\end{minipage}
\parbox{1cm}{\beq \eeq}

where on the right hand side the full fermion propagator and
the exact boson--fermion vertex appears. 

Now the standard procedure to get a closed selfconsistency equation is 
to replace the exact vertex by the tree level vertex. There are many 
justifications for this approximation. The most important point is
that it is a consistent approximation. It respects all symmetries
of the theory. Especially the Ward identities are preserved in the case of 
a gauge theory since the vertex can be chosen in an appropriate manner. 
Further in a non--abelian gauge theory with $N_c$ colors
this approximation is identical to the leading $1/N_c$ term.
For more details see \cite{Cornwall}.

Our starting point will therefore be the gap equation
\bea\label{e023}
iS^{-1}(p)-p\!\!/=
           \frac{1}{i}\int\frac{d^4k}{(2\pi)^4}
           \Gamma^a S(k)\Gamma_a D(p-k)   
\eea
where $S(p)=i(A(p^2)p\!\!/-B(p^2))^{-1}$ is the fermion propagator,
$D(q)$ the propagator of the interacting boson and $\Gamma_a$ the 
boson--fermion vertex proportional to some coupling $g$.
The index $a$ runs over Lorentz and other group indices. 
The wavefunction $A(p^2)$ does not play any important role in the game. 
Hence we set $A(p^2)\equiv 1$ for simplicity\footnote{One can also get 
rid of $A(p^2)$ by using the Landau gauge, if $D(q)$ is a gauge boson 
propagator.}.
We get:
\bea\label{e024}
B(p^2)=i\int\frac{d^4k}{(2\pi)^4}\,
       \frac{\Gamma^a\Gamma_a B(k^2)}{k^2-B(k^2)^2}\,\frac{1}{(p-k)^2} 
\eea
or for Euclidean momenta:
\bea\label{e025}
B(-p^2)=-C\int\frac{d^4k}{\pi^2}\,
       \frac{B(-k^2)}{k^2+B(-k^2)^2}\,\frac{1}{(p-k)^2} 
\eea
where $C=\Gamma^a\Gamma_a/(4\pi)^2$.
Since $\Box\frac{1}{q^2}=-4\pi^2\delta^4\!(q)$, we get:
\bea
  {1\over 4}\Box B(-p^2) \equiv
  \frac{1}{p^2}\frac{d}{dp^2}\left(p^4\frac{d}{dp^2}B(-p^2)\right) = 
  C\frac{B(-p^2)}{p^2+B(-p^2)^2}.
\eea
There are two boundary conditions \cite{Miransky}
\beq
  \[ B(-p^2)-\frac{d}{dp^2}\( p^2 B(-p^2) \) \]_{p^2=0}=0 \;
  \;\;\; \mbox{and} \;\;\; \left.\frac{d}{dp^2}\( p^2 B(-p^2) \)\right|_
  {p^2=\Lambda^2}=0
\eeq
($\Lambda$ is the momentum space cutoff.)
which are fulfilled, if $B(0)$ is finite and $C$ is above the critical 
value $1/4$ \cite{Maskawa}. 

We simplify and rotate back to Minkowski space:
\bea
\left( \frac{d}{dp^2}\right)^2(p^2B(p^2))=C\frac{B(p^2)}{p^2-B(p^2)^2}
\label{minkowski}
\eea
This scale independent differential equation can be written in an 
autonomous form replacing $p^2=e^t$ and $B(p^2)=e^{t/2}y(t)$:
\bea \label{y2p}
\ddot{y}+2\dot{y}+\frac{3}{4}y-\frac{Cy}{1-y^2}=0.
\eea
It looks like an equation of motion of a classical mass point moving with 
friction $2\dot y$ in a potential
\bea
U(y)= \frac{3}{8}y^2+\frac{C}{2}\ln|y^2-1|.
\eea
A finite\footnote{$B(0)$ can be scaled by shifting $t$ due to scale
invariance.
The solution is therefore independent of any boundary conditions.} 
$B(0)$ corresponds to $y(-\infty)=\infty$ and from $B(p^2)\to 0$ we
find $y(t)\to 0$, which should happen far beyond the mass generation scale
i.e. in the limit $p\to\infty$.
Hence the system must pass the peak of the potential at $y=1$. For $C<0$, 
as it is the case in a vector theory, $U(1)=+\infty$ and $y$ will never 
cross, so that $y>1$ and $B(p^2)^2>p^2$ for all $p^2$. Thus the fermion 
has no mass pole, which was first mentioned by Fukuda and Kugo 
\cite{Fukuda}. This problem is
connected to the infrared problem of QED, but it is hard to solve in the 
Schwinger--Dyson picture in an honest way. This problem is avoided when
replacing $B(p^2)^2$ in the denominator of eq.(\ref{minkowski}) by a
constant. 
Unfortunately this very popular approximation \cite{Miransky} destroys 
much of the nice structure, which we will explore in the sequel.

For a positive $C$, e.g. in an axial vector theory, 
one does not run into difficulties and we will concentrate on this case.
Our calculation can also be performed for negative
$C$, but it needs more effort to deal with the infrared problem.

\begin{figure}[ht]
\hspace*{0.25cm}
\begin{minipage}{7.5cm}
\begin{center}
\begin{picture}(200,200)
  \epsfysize=8.0cm
  \epsffile[600 100 1100 600]{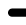}
\end{picture}
\end{center}
\vspace*{-8cm}
\begin{center}
\begin{picture}(200,200)
  \put(-10,95){$\Im(u)$}
  \put(100,-10){$\Re(u)$}
\end{picture}
\end{center}
\begin{fig} 
  Phaseportrait of the vectorfield in the complex plane for $C=0.5$.
  \label{vectorfield}
\end{fig}
\end{minipage}
\hspace*{0.4cm}
\begin{minipage}{7.5cm}
\begin{center}
\begin{picture}(200,200)
  \epsfysize=8.0cm
  \epsffile[600 100 1100 600]{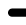}
\end{picture}
\end{center}
\vspace*{-8cm}
\begin{center}
\begin{picture}(200,200)
  \put(-10,95){$\Im(u)$}
  \put(100,-10){$\Re(u)$}
\end{picture}
\end{center}
\begin{fig} 
  Solution for $C=0.5$.  \\ \hspace*{2cm}
  \label{solution}
\end{fig}
\end{minipage}
\hspace*{0.25cm}
\end{figure}

\begin{figure}[htb]
\begin{center}
\begin{minipage}{9cm}
\begin{center}
\begin{picture}(250,180)
  \epsfysize=8.0cm
  \epsffile[600 100 1100 600]{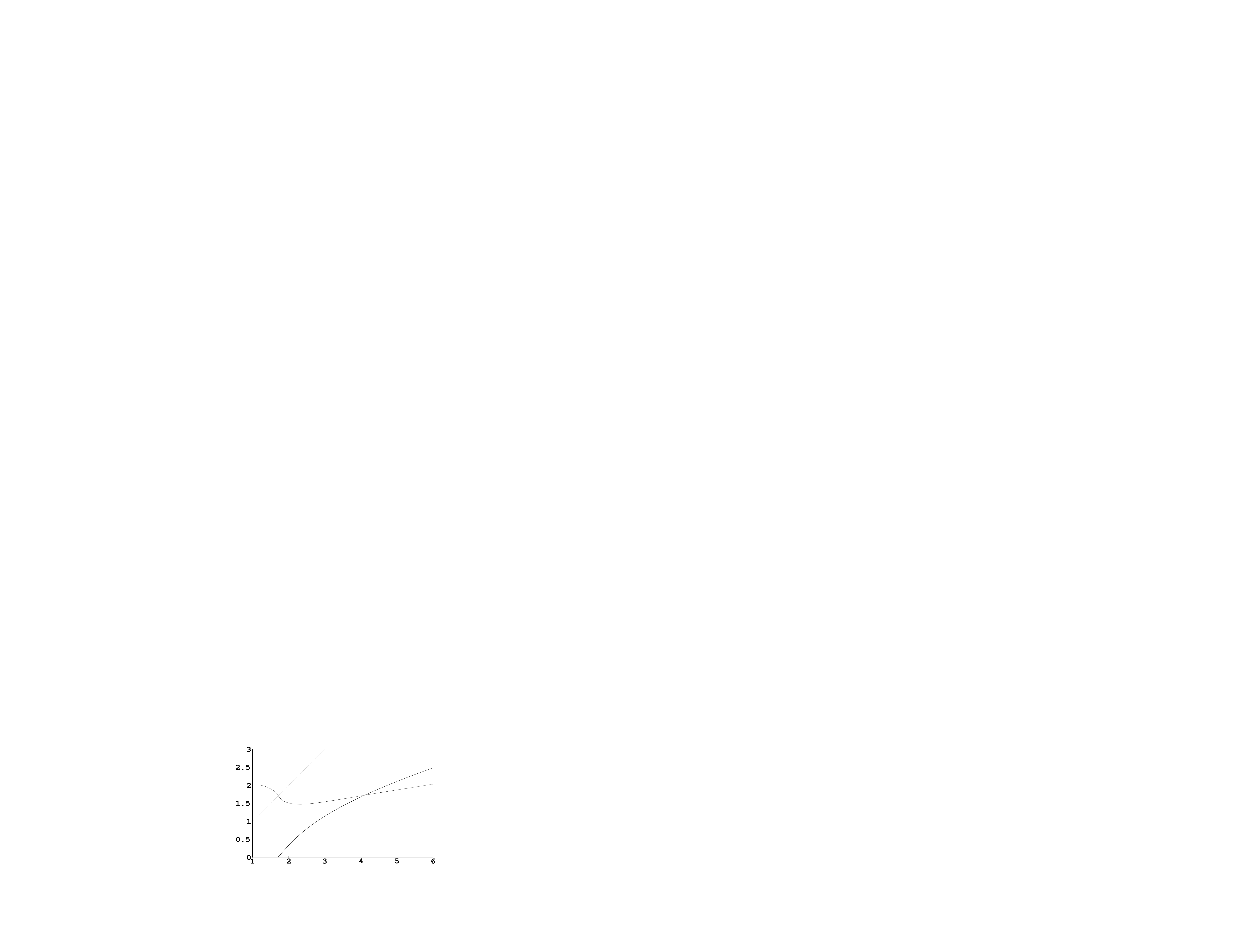}
\end{picture}
\end{center}
\vspace*{-8cm}
\begin{center}
\begin{picture}(250,180)
  \put(70,90){$\Re(B)$}
  \put(70,30){$\Im(B)$}
  \put(105,-10){$p$}
\end{picture}
\end{center}
\begin{fig}
  $B(p^2)$ for $C=0.5$.
  \label{bp2}
\end{fig}
\end{minipage}
\end{center}
\end{figure}

The line $y=1$ corresponds to the line $p^2-B(p^2)^2=0$. All $t$'s with 
$y(t)=1$ are fermion poles which we are interested in. We therefore define
$u:=y-1$ to discuss the pole behaviour at the origin. We get:
\bea\label{e8}
\ddot{u}+2\dot{u}+\frac{3}{4}(u+1)+\frac{C}{2}\left(\frac{1}{u}+
\frac{1}{u+2}\right)=0
\eea
The force vectorfield of the corresponding mechanical problem is shown in
fig.\ref{vectorfield} and a typical solution of this equation is plotted 
in fig.\ref{solution} and fig.\ref{bp2}. The solution for $u$ starts at 
real infinity. At the origin ($u=0$) the fermion pole provides a kick 
and leads to an imaginary part. For $t\to\infty$ $u$ tends to the fixed 
point -1, which corresponds to $B(p^2)\to 0$. We only get one fermion 
pole. 

Away from $u=0$ the solution follows roughly
the vectorfield in an adiabatic way indicating that the acceleration 
$\ddot{u}$ is small. Neglecting $\ddot u$, an analytic solution of
eq.(\ref{e8}) can be obtained.
The behaviour near the pole can also be studied
analytically. Near $u=0$ the potential gets large and negative.
The differential equation (\ref{e8}) is dominated by the acceleration
$\ddot u$ and the force $1/u$. Near the pole the solution is therefore
described by the equation $\ddot u+C/2u=0$ which can be solved
exactly in terms of the error function. Starting at positive $u_0$ 
the curve reaches $u=0$. Beyond this point the solution acquires
a positive/negative imaginary part if we choose $u_0\pm i\varepsilon$
as our starting point. Note that there is no kink at the origin.
The curve bends smoothly into the complex plane. 

As shown
in this approximation we do not find any periodicity of the solution.
The curve in fig.\ref{solution} does not bend back to cross the origin
again.


\section{Periodic solutions for gap equations with a running coupling}
\label{per}

The situation changes if we introduce a running coupling constant. The
simplest way to do that is to replace the constant factor $C$ in
eq.(\ref{minkowski}) by a momentum dependent function ${\cal C}(p^2)$.
It effectively describes the solution of a Schwinger--Dyson equation for
the vertex function. But a running coupling destroys the scale invariance
of our differential equation. Two effects must be separated:
One is a smooth logarithmic momentum dependence connected with the 
anomalous dimension of the coupling, which we will ignore until the 
discussion of the number of generations at the end of this section.
The remaining non-anomalous coupling ${\cal C}$ must be a function of 
dimensionless quantities and usually describe mass threshold effects.
In our case it must therefore depend non-trivially on $p^2/B(p^2)^2$
caused by the dynamically generated masses. 
Going along the steps in section \ref{gap} we end up
with
\bea\label{e31}
\ddot{u}+2\dot{u}+\frac{3}{4}(u+1)+\frac{{\cal C}(u)}{2}\left(\frac{1}{u}+
\frac{1}{u+2}\right)=0
\eea
which can be parametrized by a Taylor expansion up to the quadratic term:
\bea  \label{dgl}
\ddot{u}+2\dot{u}+\frac{3}{4}(u+1)+\frac{{\cal C}(0)}{2}\left(\frac{1}{u}+
\frac{1}{u+2}\right)+\alpha+\beta u+\gamma u^2=0
\eea
Now for a wide range of parameters one gets periodic solutions as in
fig.\ref{solution2}, where we have chosen ${\cal C}(0)=1.5$, $\alpha=-1.5$,
$\beta=-1.0$ and $\gamma=1.0$. The insertion of this solution into the gap
equation for the vertex function would determine in turn the coupling
${\cal C}(u)$ which should be consistent with our ansatz. 
This is a difficult task, which will not be addressed in this work.
Instead we directly want to study the phenomenological consequences
of the solutions.

\begin{figure}[htb]
\begin{center}
\begin{minipage}{10cm}
\begin{center}
\begin{picture}(200,200)
  \epsfysize=8.0cm
  \epsffile[600 100 1100 600]{plot5.ps}
\end{picture}
\end{center}
\vspace*{-8cm}
\begin{center}
\begin{picture}(200,200)
  \put(-10,95){$\Im(u)$}
  \put(100,-10){$\Re(u)$}
  \put(147,117){$\scriptstyle u_F$}
  \put(147,63){$\scriptstyle u_F^*$}
\end{picture}
\end{center}
\begin{fig}
  Solution for ${\cal C}(0)=1.5$, $\alpha=-1.5$, $\beta=-1.0$ and
$\gamma=1.0$.
  There are two fixed points $u_F^{(*)}=0.646\pm 0.617i$.
  \label{solution2}
\end{fig}
\end{minipage}
\end{center}
\end{figure}

\begin{figure}[htb]
\begin{center}
\begin{minipage}{10cm}
\begin{center}
\begin{picture}(250,200)
\epsfysize=8.0cm
\epsffile[600 100 1100 600]{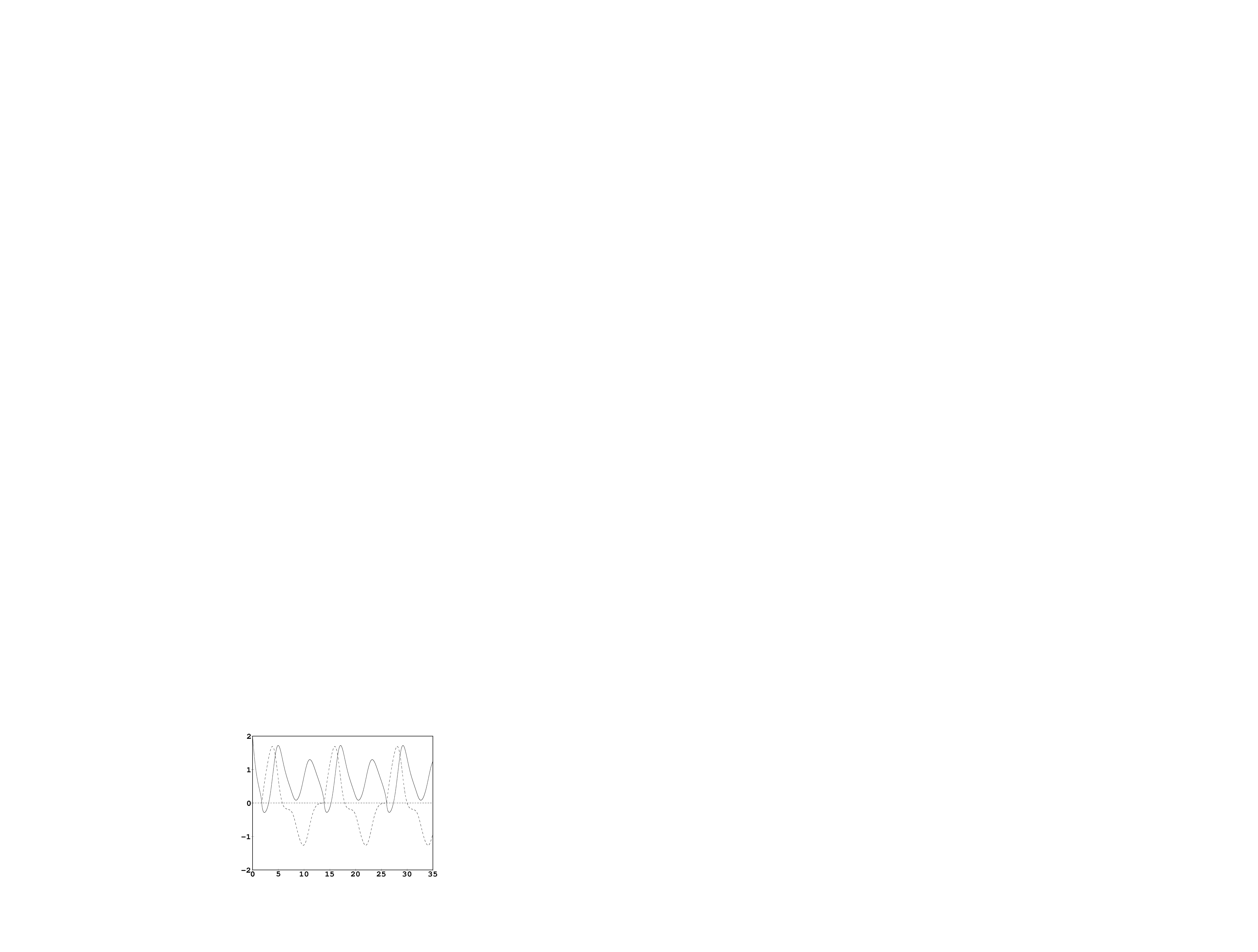}
\end{picture}
\end{center}
\vspace*{-8cm}
\begin{center}
\begin{picture}(250,200)
\put(50,30){$\Im(u)$}
\put(50,160){$\Re(u)$}
\put(105,-10){$\log p^2$}
\end{picture}
\end{center}
\begin{fig}
  Real and imaginary part for ${\cal C}(0)=1.5$, $\alpha=-1.5$, 
  $\beta=-1.0$ and $\gamma=1.0$.
  \label{imag}
\end{fig}
\end{minipage}
\end{center}
\end{figure}

The behaviour near $u=0$ is unchanged and the analysis of the last section 
can be adopted. But for large $u$ the solution
now bends back to the real positive axis allowing periodic solutions.
In each cycle $u=0$ is passed from the right to the left in the same way.
Therefore all poles have the same residue. Positivity of all residues
guarantees that all poles correspond to physical particles (not ghosts).
The poles of the propagator occur at $u(0)=u(T)=u(2T)=\ldots=0$, where
$T$ is the period of the solution. For the example given above $T=12.1$.
From the stability condition of the trajectory around the fixed point
$u_F$ of the differential equation (\ref{dgl}) one can roughly 
estimate\footnote{This can be calculated by expanding eq.(\ref{dgl}) around
$u_F$ and demanding a stable circle as the orbit of the system around that
fixed point.}:
\beq \label{approx}
  T\approx 2\pi L\cdot \left|\frac{\Re(u_F)}{\Im(u_F)}\right|
     \(1+\sqrt{1-2\gamma \Re(u_F)|u_F|^2} \)^{-1}
  \approx  2\pi L
\eeq
for $\Im(u_F)\approx \Re(u_F)$ and generic values for $\gamma$. 
$L$ is the number of loops of the trajectory around the
fixed points. In fig.\ref{solution2} there are 2 fixed points inside the
trajectory and hence $L=2$.
Curves surrounding only one fixed point are possible, too (see
fig.\ref{sim2}). 
However the coupling ${\cal C}(u)$ needs a soft fine tuning in that case.

Now we get infinitely many
elementary fermionic excitations with an exponential mass spectrum
\beq
  m_n = m_0\,e^{nT/2},\quad n\in \mbox{\sf Z\hspace*{-1.8mm}Z}
\eeq
Scale invariance of eq.(\ref{e023}) by an arbitrary factor $\lambda$
is broken dynamically down to a discrete subgroup generated by
a scaling with $e^{T/2}$, which transforms a state of mass $m_n$ 
into a state of mass $m_{n+1}$.

One can try to write down an effective action with an elementary
field for each $n$. Due to the fact that the residue of all poles is
the same, the fermions will be coupled in a univeral manner to the
gauge boson. Absence of flavor changing currents and the stability of the
spectrum will be discussed in section \ref{eff}.
Many features of the family structure of the
standard model have so far be reproduced in a qualitative manner. Even
the large mass ratios of the order of
\beq
   m_{n+1}/m_n = e^{T/2} \approx 16 \ldots 300
\eeq
with \cite{Koide}
\beq
  T_{\mu-\tau}=5.6,\;\; T_{d-s}=6.0,\;\; T_{s-b}=7.1,\;\; T_{e-\mu}=10.7,
  \;\; T_{u-c}=11.2,\;\; T_{c-t}=11.3
\eeq
can be understood in our model.

The down--type quarks $d-s-b$ can be associated with $L=1$:
\beq
  T_{d-s}\approx T_{s-b}\approx 2\pi,
\eeq
the up--type quarks $u-c-t$ with $L=2$:
\beq
  T_{u-c}\approx T_{c-t}\approx 4\pi.
\eeq
The mass ratios between the different generations can therefore be reduced 
to a number of natural size.\footnote{The leptons seem to have a mixing 
between both periods, which could happen by a variation of the 
corresponding coupling.} 
Although the approximation (\ref{approx}) is very rough
and depends on the special parametrization, the factor 2 between the
two different periods $2\pi$ and $4\pi$ is purely topological and does not
depend on the special ansatz and parametrization. 

Why are there only three generations? We want to give some speculations
why there are only a finite number of generations: The periodic solution
cannot be the whole story for a theory with non--vanishing beta function.
For a negative beta function\footnote{For $\beta>0$ the discussion is the
same except for interchanging large and small distances.}
asymptotic freedom has to be incorporated.
At small distances the coupling gets small and perturbation theory should
be valid. No dynamical mass generation is expected and the periodic
solution should converge into the one loop expression. For large distances
below the quantum scale $\Lambda$ of the theory, where the coupling
gets large, or below a possible current mass of the fermion the theory
is no longer approximately scale invariant. But scale invariance
was essential for arriving at the translation invariant differential
equation (\ref{e31}) and the periodic solution. Periodicity of the
solution is therefore only expected in a region of intermediate coupling.
This region has finite size (for $\beta\neq 0$) and thus the number of 
generations is expected to be finite. Within this approach the value $3$ 
for the number of generations is just a dynamical accident.

A crucial point, which we have not yet addressed, is the question for
the underlying interaction responsible for that dynamical symmetry breaking
scenario.
To explain it by merely Standard Model interactions seems to be impossible.
On the other side a new interaction is hard to introduce since the new 
bosons should be nearly massless. The gauge group must be hidden in some 
way to forbid vector boson interactions at tree level and to admit only 
self energy and vertex corrections. 
For the future an explicit model should be constructed.


\section{Multipol propagator versus dynamical mass function} \label{mul}  

How can one explore whether some particles with the same quantum numbers
really come from different propagators or belong to one propagator with 
several poles? One would naively suppose that there must be some essential
differences. 
Starting with an exponential fermionic mass spectrum 
and cutting the spectrum as described in section \ref{per}, a theory of 
$N_f$ free fermions remains.
The sum of their propagators can be interpreted as one propagator with
a mass function $B(p^2)$:
\beq
 \frac{i}{p\!\!/-B(p^2)} = \sum_{k=1}^{N_f}
\frac{i}{p\!\!/-m_k}\;,\;\;\;
 \mbox{where} \;\; m_k:=m_1 e^{\alpha (k-1)} 
\eeq
As explained in the introduction for $N_f>1$, $B(p^2)$ is no longer a well
behaved selfenergy, but has poles. To avoid them each
fermion self energy has to acquire an imaginary part as it appears e.g. 
in an one--loop calculation, if we admit an interaction:
\beq
 \frac{i}{p\!\!/-m_k}\to \frac{i}{p\!\!/-m_k}+\frac{i}{p\!\!/-m_k}
 (-i\Sigma_k) \frac{i}{p\!\!/-m_k}
\eeq
where 
\beq
 \Sigma_k = c\cdot i\cdot \theta(p^2-m_k^2)\frac{p^2-m_k^2}{p^2}\cdot m_k
\eeq
with some constant $c$.
The additional terms can be interpreted as some kind of continuum produced
by that interaction.
Extracting the $p\!\!/$--independent part one can calculate the function
$u(t)$ from $B(p^2)=e^{t/2}(u(t)+1)$. It is plotted in fig.\ref{sim}.
\begin{figure}[ht]
\hspace*{0.25cm}
\begin{minipage}{7.5cm}
\begin{center}
\begin{picture}(200,200)
  \epsfysize=8.0cm
  \epsffile[600 100 1100 600]{plot7.ps}
\end{picture}
\end{center}
\vspace*{-8cm}
\begin{center}
\begin{picture}(200,200)
  \put(-10,95){$\Im(u)$}
  \put(100,-10){$\Re(u)$}
\end{picture}
\end{center}
\begin{fig} 
  u from the sum of several independent propagators.
  \label{sim}
\end{fig}
\end{minipage}
\hspace*{0.4cm}
\begin{minipage}{7.5cm}
\begin{center}
\begin{picture}(200,200)
  \epsfysize=8.0cm
  \epsffile[600 100 1100 600]{plot8.ps}
\end{picture}
\end{center}
\vspace*{-8cm}
\begin{center}
\begin{picture}(200,200)
  \put(-10,95){$\Im(u)$}
  \put(100,-10){$\Re(u)$}
\end{picture}
\end{center}
\begin{fig} 
  Solution of the gap equation with $L=1$.
  \label{sim2}
\end{fig}
\end{minipage}
\hspace*{0.25cm}
\end{figure}
One obviously gets a periodic structure similar to fig.\ref{sim2}.
Our differential equation (\ref{dgl}) with
${\cal C}(0)=1.5$, $\alpha=-1.5$, $\beta=-0.453$ and $\gamma=1.0$ yields
that solution, where $L=1$.
One should not expect exact coincidence because we are comparing an one 
loop calculation with a solution of a much more complicated gap equation.

Hence our solution can effectively be described by the sum of 
independent propagators.
A gauge field coupled to {\it the} fermion field seems to allow the decay 
of our states by going to different on--shell limits for an incoming and
the corresponding outcoming fermion line. This is {\it not} true and we 
will prove it in the next section.


\section{Effective Lagrangians and the stability of the 
         mass spectrum}
\label{eff}

Starting from a scale invariant theory of one quark flavor 
interacting with some boson field we saw how to obtain multiple
flavors with an exponential mass hierarchy. This has been achieved
by analyzing gap equations for the fermion propagator $S$. As shown in the
last section the solution takes the form 
\beq\label{e19}
  S(p)=\sum_{k=1}^{N_f}{i\over p\!\!/-m_k} \quad+\quad\mbox{\it continuum}
\eeq  
We ignore the continuum for a while, which describes some residual 
interaction.  The effective Lagrangian 
\beq\label{e20}
  {\cal L}_{1,free}^{eff} =
  i\bar\psi S_0^{-1}\psi \quad,\quad 
  S_0=\sum_{k=1}^{N_f}{i\over i\partial\!\!\!/-m_k}
\eeq
reproduces $S$ (without continuum) and hence
has the same particle spectrum. The Lagrangian is non-local due to the
non-local kernel $S_0^{-1}$, but despite this it describes a free
theory of pointlike particles, because it is quadratic in the fermion 
fields.
The more familiar local Lagrangian
\beq\label{e21}
  {\cal L}_{N_f,free} =
  \sum_{k=1}^{N_f}\bar\psi_k(i\partial\!\!\!/-m_k)\psi_k
\eeq
containing explicitly $N_f$ fermion field operators $\psi_k$,
one for each single particle state,
describes also the same physics. One might interpret 
(\ref{e21}) as an effective Lagrangian of (\ref{e20}) which is 
itself (the free part of) an effective Lagrangian of the original 
theory.

There are two types of interactions which should be incorporated 
into eq.(\ref{e20}) now. One is some residual interaction
of the mass--generating boson studied in the gap equation and coded
in the continuum contribution to eq.(\ref{e19}). The other are interactions
with other (standard model) gauge bosons. E.g. electromagnetism, when
present in the original theory, must also emerge in some way in the 
effective theory. The principles of gauge invariance
and minimal substitution $(\partial\!\!\!/ \to D\!\!\!\!/)$ 
dictates the form of the interaction in both cases:  
\beq\label{e22}
  {\cal L}_1^{eff} =
  -{1\over 4}F_{\mu\nu}F^{\mu\nu} + i\bar\psi S^{-1}\psi \quad,\quad
  S=\sum_{k=1}^{N_f}{i\over iD\!\!\!\!/-m_k}     \quad,\quad
  D\!\!\!\!/=\partial\!\!\!/+igA\!\!\!/
\eeq
$D\!\!\!\!/$ is the covariant derivative depending on the gauge field 
$A_\mu$ and $F_{\mu\nu}$ is its field strength.
The scattering of $n$ quarks can be obtained as usual from the 
$2n$-point function. Integrating out {\it the}\footnote{
Note that there is still only one fermion field.} fermion field
in the background of the gauge field $A_\mu$
we get the path integral representation for the $2n$-point 
function\footnote{
To get familiar with the operator notation and the background techniques
one should consult \cite{NSVZ,Itzykson}.}
\beq\label{e23}
  \langle 0|\psi(x_1)\bar\psi(y_1)\ldots
            \psi(x_n)\bar\psi(y_n)|0\rangle =
\eeq
$$
  = \int\! \D A_{\mu}\;e^{-i\int\!d^4\!x{1\over 4}F^2(x)}
  \mbox{Det}S^{-1} \prod_{l=1}^n
  \langle x_l|\sum_k{i\over iD\!\!\!\!/-m_k}|y_l\rangle
  \quad+\quad\mbox{\it crossed terms}
$$
For each contraction of a $\psi$ with a $\bar\psi$, we get a propagator
$\langle x|S|y\rangle$ in coordinate representation in
the background of the gauge field $A_{\mu}$. $\mbox{Det}S^{-1}$ is
the functional determinant of the kernel.
To get e.g. the propagator (\ref{e19}) one has to use eq.(\ref{e23})
with $n=1$, which is just the integration of the propagator (\ref{e22}) 
over the quantum fluctuations of the gauge field. 

It is instructive to compare eq.(\ref{e23}) to the gauged version
of eq.(\ref{e21}), the standard model of $N_f$ fermions coupled to a
gauge field
\beq\label{e24}
  {\cal L}_{N_f} =
  -{1\over 4}F_{\mu\nu}F^{\mu\nu} +
  \sum_{k=1}^{N_f}\bar\psi_k(iD\!\!\!\!/-m_k)\psi_k
\eeq
with
\beq\label{e25}
   \langle 0|\psi_{k_1}(x_1)\bar\psi_{k_1}(y_1)\ldots
            \psi_{k_n}(x_n)\bar\psi_{k_n}(y_n)|0\rangle =
\eeq
$$
  = \int\! \D A_{\mu}\;e^{-i\int\!d^4\!x{1\over 4}F^2(x)}
  \prod_{k=1}^{N_f}\mbox{Det}(iD\!\!\!\!/-m_k) \prod_{l=1}^n
  \langle x_l|{i\over iD\!\!\!\!/-m_{k_l}}|y_l\rangle
  \quad+\quad\mbox{\it crossed terms}
$$
The only difference between eq.(\ref{e23}) and eq.(\ref{e25}) are the
functional determinants. At tree level, or more general in quenched 
approximation, both expressions coincide\footnote{
Taking the on--shell limit selects the appropriate flavor from the
sum on the right hand side.}:
\beq\label{e26}
   \langle 0|\psi(x_1)\bar\psi(y_1)\ldots
            \psi(x_n)\bar\psi(y_n)|0\rangle =
   \sum_{k_1\ldots k_n}
   \langle 0|\psi_{k_1}(x_1)\bar\psi_{k_1}(y_1)\ldots
            \psi_{k_n}(x_n)\bar\psi_{k_n}(y_n)|0\rangle 
\eeq
In summary we can say that our model of mass generation can be described
by the effective action ${\cal L}_1^{eff}$ which reduces to 
${\cal L}_{N_f}$
in quenched approximation. Dynamical quark loops incorporated by
the functional determinant lead to effective 
non--local boson vertices as in the standard theory ${\cal L}_{N_f}$, but
differ from them in magnitude.

One immediate consequence of the discussion above is the absence of flavor 
changing currents. 
For ${\cal L}_{N_f}$ this is obvious and therefore it must be true for
${\cal L}_1^{eff}$ at tree level due to the equivalence proven above.
Further, the inclusion of dynamical quark loops has no influence
on the flavor structure of the external fermions.
This is consistent with the reality of the masses $m_n$, which also 
indicates the stability of the particles (but in a much less convincing 
way).

Feynman rules for the Lagrangian ${\cal L}^1_{eff}$ are derived in the
appendix. They differ from the standard feynman diagrams of 
${\cal L}_{N_f}$ only by one additional flavor changing graph

\begin{minipage}{14.4cm}
\begin{center}
\begin{picture}(230,20)(0,5)
\Text(10,12)[c]{$k$}
\ArrowLine(20,10)(45,10)        \ArrowLine(45,10)(70,10)
\Line(42,7)(48,13)              \Line(42,13)(48,7)
\Text(80,12)[c]{$k'$}
\Text(95,10)[c]{=}
\Text(110,12)[l]{$-S_0^{-1}$}
\end{picture}
\end{center}
\end{minipage}
\parbox{1.5cm}{\beq \label{s0} \eeq}

connecting flavor $k$ with $k'$, which is shown to be allowed only
in closed fermion loops. This provides a perturbative proof of the 
stability of the mass spectrum.
In the next section we will show,
that everything said above applies to electromagnetism,
weak and strong interaction, where $\psi$ as well as each $\psi_k$ 
have to be interpreted as a single fermion, a weak doublet or 
a color triplet respectively.


\section{More realistic models}        \label{real}

In this section we want to scetch what is changed when going from
the toy model with one flavor per generation to more realistic
models with two flavors per generation. We consider an 
$U(1)\!\times\! SU(2)$ gauge theory, which is not the Standard Model but 
rather a strongly coupled hidden gauge theory, with abelian coupling $e$,
vector coupling $g_V$ and axial coupling $g_A$
\beq
  {\cal L}_{int} = e\bar{\psi^m}A\!\!\!/\psi^m + 
  \bar{\psi^m}(g_V\!-\!g_A\gamma_5)W\!\!\!\!\!/_a\tau^a_{mn}\psi^n
  \quad,\quad m,n=1,2 \quad,\quad a=1,2,3
\eeq
where summation over isospin indices $m$, $n$ and $a$ is understood.
$\psi$ is now an isospin doublet combining an up-type quark with a
down-type quark $({u\atop d})$ (or a lepton with its neutrino 
$({\nu_l\atop l})$).
The propagators of both fermions can be combined into an isodoublet
matrix like for the Dirac-operator. 

For an isospin symmetric ansatz $S^{mn}=i\delta^{mn}/(p\!\!/-B)$,
the discussion in the previous chapters achieves only minor modifications.
The gap equation (\ref{e024}) can be derived, with $\Gamma_a$ replaced by
$\Gamma_\mu = -ie\gamma_\mu$ and 
$\Gamma_\mu^a=-i(g_V\!-\!g_A\gamma_5)\gamma_\mu\tau^a$.
We obtain the equation (\ref{y2p}) with
\beq
  C = (\Gamma_\mu\Gamma^\mu+\Gamma_\mu^a\Gamma^\mu_a)/(4\pi)^2 =
  -{e^2\over 4\pi^2} - 3\!\cdot\!{g_V^2-g_A^2\over 4\pi^2}.
\eeq
Replacing the coupling constant $C$ by a running couping ${\cal C}(u)$
we get periodic solutions and hence an exponential mass spectrum,
where the up-type quarks are degenerate in mass with the down-type 
quarks. There is no isospin breaking, of course, because we have chosen
a symmetric ansatz. The discussion of section \ref{eff} remains valid,
if we interpret $\psi$ and each $\psi_k$ as an isodoublet and use
$D\!\!\!\!/=\partial\!\!\!/+ieA\!\!\!/+
            i(g_V\!-\!g_A\gamma_5)\tau^aW\!\!\!\!\!/_a$.
Especially there are no FCNC and the CKM matrix is identical to $1\!\!1$.

An isovector ansatz $S=i/(p\!\!/\!-\!B_a\tau^a)$
is also a solution of eq. (\ref{e23}). If we choose the orientation of
$B_a(p^2)$ constant in isospace, say $B_a(p^2)=(0, 0, B(p^2))$ we
arrive again at eq. (\ref{e025}) with $C$ replaced by\footnotemark
\beq
  \widetilde C = (\Gamma_\mu\tau^b\Gamma^\mu+\Gamma_\mu^a\tau^b
  \Gamma^\mu_a)/
             [(4\pi)^2\tau^b] =
  -{e^2\over 4\pi^2} + {g_V^2-g_A^2\over 4\pi^2}
\eeq
\footnotetext{Both signs of $C$ and $\widetilde C$ can be realized as 
anticipated in section \ref{gap}}
Again for a running coupling we get an isospin degenerate exponential 
mass spectrum, but now the up- and down-type masses differ in an 
irrelevant sign $(m_u=-m_d)$.

The most general (parity even) ansatz
\beq 
  S = {i\over p\!\!/-B_0-B_a\tau^a}
\eeq
leads to a system of coupled differential equations
\bea\label{e57}
   \ddot y_0+2\dot y_0+{3\over 4}y_0+{C\over 2}
   \left({y_0-1\over(y_0\!-\!1)^2-\vec y^2} + 
         {y_0+1\over(y_0\!+\!1)^2-\vec y^2} \right) &=& 0 \\
   \ddot y_a+2\dot y_a+{3\over 4}y_a-{\widetilde C\over 2}
   \left({y_a\over(y_0\!-\!1)^2-\vec y^2} + 
         {y_a\over(y_0\!+\!1)^2-\vec y^2} \right) &=& 0 \nonumber
\eea
with $\vec y^2=y_ay^a$ and $C$ and $\widetilde C$ are defined above.
The physical masses occur at the poles $(y_0\pm1)^2=\vec y^2$.
One should now study general solutions with running coupling and select
those with the lowest vacuum energy. These solutions might provide an
interesting isospin breaking pattern with an interesting CKM matrix in
addition to the exponential mass spectrum.  

When studying more general gauge groups, 
e.g.\ a GUT theory combining all fermions of one generation 
into a multiplet, eq.(\ref{e57}) remains
valid. Only the definition of $C$ and
$\widetilde C$ and the number of components of $y_a$ changes. 


\section{Conclusion}

In this paper we gave a new approach to understand the Standard Model 
fermion mass spectrum. Since a possible horizontal symmetry must be broken
drastically, we have instead investigated an improved gap equation with 
periodic solutions. This equation is triggered by a new strongly coupled 
boson and yield a fermion propagator with a series of poles.
The scale invariant structure of the system automatically leads to the 
desired exponential particle spectrum. All particles have the same residue 
and couple to all bosons with the same strength.
The mass ratios between particles of successive generations turn out to 
be 
\beq
     \frac{m_{n+1}}{m_n}=e^{L\pi{\cal O}(1)}
\eeq
with $L=1$ or 2 and can therefore be attributed to numbers of natural
size. We found a good agreement on logarithmic accuracy with the spectrum 
of the up and down--type quarks of the Standard Model.

The stability of the excited fermions, which is the most difficult problem,
is solved in the effective Lagrangian formalism. Gauge invariance dictates
the structure of the effective theory which inevitably forbids flavor
changing neutral transitions.

Because of this success the model seems to be a good candidate for the 
mass generation and especially to explain the exponential structure of the
spectrum. A correct inclusion of the anomalous dimension of the coupling
should lead to a finite number of generations as discussed in section 
\ref{per}. The next step will be to specify the new interaction and to
determine the detailed structure of the mass generation.
Then the effective Lagrangian can be extracted more rigorously from that
specific model. 

For the future it would also be of great interest to study loop 
corrections, which admit virtual flavor changing currents and hence lead 
to deviations from standard model predictions.
Finally a complete model would offer the possibility to calculate the
CKM matrix and therefore to test the theory accurately.


\begin{appendix}
\vspace{1.5cm}{\Large \bf Appendix} \vspace{-0.5cm}
\renewcommand{\theequation}{\Alph{section}.\arabic{equation}}
\setcounter{equation}{0}

\section{Feynman rules of the effective model}

To determine the Feynman rules of the effective model discussed in section
\ref{eff} we start with the partition function
\beq
   Z[\bar{\eta},\eta,j]=\int\!\D A_{\mu}\D\psi\D\bar{\psi}\,
   e^{\textstyle i\int\!d^4\!x({\cal L}+\bar{\eta}\psi+\bar{\psi}\eta+
   j^{\mu}A_{\mu})}
\eeq
where
\beq
  {\cal L} =
  -{1\over 4}F_{\mu\nu}F^{\mu\nu} + i\bar\psi S^{-1}\psi \quad,\quad
  S=\sum_{k=1}^{N_f}{i\over iD\!\!\!\!/-m_k}     \quad,\quad
  D\!\!\!\!/=\partial\!\!\!/+igA\!\!\!/.
\eeq
$S^{-1}$ can be expanded using $\Delta_k:=\frac{\textstyle 1}{\textstyle 
i\slash{\partial}-m_k}$ and $S_0:=i\sum\limits_{k=1}^{N_f}\Delta_k$:
\beq
  S^{-1} = S_0^{-1}+V S_0^{-1}
\eeq 
with
\beq \label{v}
  V=\[ 1+iS_0^{-1}\sum_{k=1}^{N_f}\( g\Delta_k\slash{A}\Delta_k +
  g^2\Delta_k\slash{A}\Delta_k\slash{A}\Delta_k+\ldots\)\]^{-1}-1.
\eeq
Integrating out the fermion field we get
\beq
   Z[\bar{\eta},\eta,j]\propto\int\!\D A_{\mu}\;
   e^{\textstyle i\int\!d^4\!x \(
   -{1\over 4}F_{\mu\nu}F^{\mu\nu}+j^{\mu}A_{\mu}\)} \; 
   Z[\bar{\eta},\eta;A]
\eeq
with
\beq
   Z[\bar{\eta},\eta;A] = 
   e^{\textstyle -i\int\!d^4\!x \frac{\partial}{\partial\eta}
   iV S_0^{-1} \frac{\partial}{\partial\bar{\eta}} }\;
   e^{\textstyle i\int\!d^4\!x \,\bar{\eta}iS_0\eta }.
\eeq
Terms, which contain one open fermion line and an arbitrary number of 
boson lines, can be extracted by an expansion in $V$. The $n$ 
boson--fermion vertices can be ordered in $n!$ different ways. Thus the 
factors in the exponential series are compensated by combinatorical 
factors. We therefore get a geometric series:
\bea \label{zo}
  Z_{open}[\bar{\eta},\eta;A] &=&
  -\bar{\eta}S_0\(1-V+V^2-V^3+-\ldots\)\eta  \nonumber \\
  &=& -i\sum_{k=1}^{N_f}\bar{\eta}\( \Delta_k+
  g\Delta_k\slash{A}\Delta_k +
  g^2\Delta_k\slash{A}\Delta_k\slash{A}\Delta_k+\ldots\)\eta.
  \label{sum1}
\eea
There are no flavor changing transitions because there 
occurs $\Delta_k$ for only one flavor $k$ in each term. 
This is due to the cancellation of the geometric series against the 
inversion in eq.(\ref{v}).
The series (\ref{sum1}) can be summed up to -i$\bar{\eta}S\eta$, which
is the tree level part of eq.(\ref{e23}).
 
Terms, which contain one closed fermion line and an arbitrary number of 
boson lines, are
\bea \label{zc}
  Z_{closed}[\bar{\eta},\eta;A] \!\!&=& \!\!\!\!\label{sum2}
  -i\mbox{Tr}\[\(V-\frac{1}{2}V^2+\frac{1}{3}V^3-+\ldots\)\]  \\
  &=& \!\!\!
  ig^2\mbox{Tr}\[\sum_{k=1}^{N_f}\Delta_k\slash{A}\Delta_k\slash{A}
  \Delta_k iS_0^{-1}-\frac{1}{2}\sum_{k,j=1}^{N_f}
  \Delta_k\slash{A}\Delta_kiS_0^{-1}
  \Delta_j\slash{A}\Delta_jiS_0^{-1}\]\!\!+\!{\cal O}(g^4). \nonumber
\eea
For a closed fermion loop $n$ boson--fermion vertices can be ordered in 
$(n-1)!$ different ways. Hence we find a logarithmic series, which gives
$-i\mbox{Tr}\ln (1+V)=-i\ln\mbox{Det}(1+V)$. The sum over all diagrams 
with an arbitrary number of loops is proportional to $\mbox{Det} S^{-1}$, 
which is the loop correction in eq.(\ref{e23}).
The mixed occurence of $\Delta_k$, $\Delta_j, \ldots$ in eq.(\ref{sum2}) 
shows that flavor changing transitions
appear in closed fermion lines. The transition is induced by the graph
(\ref{s0}). 

The Feynman rules can be read of eq.(\ref{zo}) and eq.(\ref{zc}).
They only differ in the loop corrections 
from the standard model (\ref{e24}). The spectrum is therefore stable.

\end{appendix}


\parskip=0ex plus 1ex minus 1ex

\end{document}